\documentclass[5p,twocolumn,number,sort&compress]{elsarticle}
\usepackage[utf8x]{inputenx}
\usepackage{microtype}
\usepackage{amsmath}
\usepackage{amsfonts}
\usepackage{graphicx}
\usepackage{color}
\begin{document}

\title{Helical bound states in the continuum of the edge states in two dimensional topological insulators}

\author{Vladimir~A. Sablikov\corref{cor1}}
\cortext[cor1]{Corresponding author}
\ead{sablikov@gmail.com}
\author{Aleksei~A. Sukhanov}
\address{V.A.~Kotel'nikov Institute of Radio Engineering and Electronics,
Russian Academy of Sciences, Fryazino, Moscow District, 141190, Russia}

\begin{abstract}
We study bound states embedded into the continuum of edge states in two-dimensional topological insulators. These states emerge in the presence of a short-range potential of a structural defect coupled to the boundary. In this case the edge states flow around the defect and have two resonances in the local density of states. The bound state in continuum (BIC) arises due to an interference of the resonances when they are close to the degeneracy. We find the condition under which the BIC appears, study the spacial distribution of the electron density, and show that the BIC has a helical structure with an electron current circulating around the defect.
\end{abstract}

 
\maketitle
\section{Introduction}
\label{Intro}

The existence of bound states embedded into the continuum of quantum states was proposed at the dawn of quantum mechanics by von Neumann and Wigner~\cite{vonNeumannWigner1929} for certain attractive potentials of very specific form. The formation of the bound state in continuum (BIC) can be considered as a result of a destructive interference of partial waves in the potential due to which the wave function vanishes at large distance. It is essential that the BICs are true eigenstates of the Hamiltonian and therefore are orthogonal to the eigenstates of the continuous spectrum. Because of a rather complicated form of the potential these states were regarded a long time as mathematical curiosities. However, further developments have produced a better understanding of the kind of potential that can create such bound states. Friedrich and Wintgen~\cite{FriedrichWintgen1985} have pointed out that the resonant states play an important role in the BIC formation and proved that a BIC could emerge as two coupled resonant states connected to a continuum are driven into degeneracy by changing a parameter governing the resonant states.  

The advances in the nanostructure technology have provided new possibilities to create the required potential and the interest to BICs has strongly increased due to their manifestations in quantum transport, optical phenomena, and potential applications for quantum information devices. Herrick~\cite{Herrick197644} and Stillinger~\cite{Stillinger1976270} have proposed a construction of the potential supporting BICs for epitaxial heterostructure superlattices. The localized states with properties very close to BICs have been realized by Capasso et al.~\cite{capasso1992observation} in semiconductor heterostructures. In recent years the BICs are widely studied also in photonic crystals and were observed experimentally~\cite{PlotnikPRL2011}. 

Theoretical studies of the occurrence of BICs in low-dimensional structures have been investigated for a variety of systems such as quantum Hall effect and narrow-wire circuits~\cite{SchultPRB1990}, systems of coupled quantum dots~\cite{RotterSadreevPRE2005,GuevaraOrellanaPRB2006,GonzalezEPL2010}, open quantum billiards~\cite{SadreevPRB2006}, trilayer graphene flakes connected to nanoribbon leads~\cite{CortesEPL2014}. A distinct mechanism of the BIC formation was recently proposed for a chiral quantum system coupled to leads with a continuum energy band~\cite{MurPetitMolinaPRB2014}.    

In the present paper we address the question of how the unique properties of topological insulators (TIs) affect the possibility of the BIC formation and come to the conclusion that at least in the case of two-dimensional (2D) TIs the emergence of BICs is facilitated by the presence of the topological order so that a BIC appears in a simple potential produced by a structural defect or impurity. 

The distinctive feature of the 2D TIs is the presence of helical edge states with massless Dirac spectrum in the gap of bulk states~\cite{Kane2013inbook}. In the edge states the electrons move along the boundary and their spin is locked to the momentum because of strong spin-orbit interaction. In recent works~\cite{SablikovPSSR2014,SablikovPRB2015} we have studied the electronic states induced by nonmagnetic defects and found that a defect with a short-range potential creates two bound states in contrast to topologically trivial insulators where only one bound state exists at such a defect. These states differ in both the pseudospin structure and the spatial distribution of the particle density. It is also interesting that they arise for any sign of the defect potential. The presence of two states is caused by the fact that in 2D TIs there are two mechanisms of the bound-state formation. One mechanism is conventional. The bound state is formed by a potential attracting the quasiparticles of one of the bands. Another mechanism is specific for TIs. It is caused by the formation of an edge state circulating around the defect similarly to the edge states near the boundary. One can say that the defect effectively creates a boundary condition for the wave function. 

When the defect is located at a finite distance from the boundary, the bound state is coupled with the edge states giving rise to the formation of the edge states flowing around the defect. These states have a continuous spectrum with resonances in the local density of states near the defect at the energy close to the bound state energy~\cite{SablikovPRB2015}.

In the present paper we study the true bound states, which form due to the interference of these resonances. These states have a square integrable wave function and the energy embedded into the continuum of the edge states flowing around the defect. Because of strong spin-orbit interaction these BICs have a helical structure.

\section{Model and equations determining the BIC}
Consider a 2D TI, which contains a defect located at a distance $y_d$ from the boundary. The defect creates a potential $V(x,y-y_d)$ localized in a small region. We use a layout in which the $x$-axis coincides with the boundary and the $y$-axis is directed into the TI. The electronic states in the 2D TI are described by the model Hamiltonian proposed by Bernevig, Hughes, and Zhang (BHZ) for HgTe/CdTe quantum wells~\cite{BHZScience2006}. This model has the $S_z$ symmetry since it does not take into account a possible spin-orbit interaction arising from the lack of the inversion symmetry. The corresponding generalization seems not to be essential at this stage and will be done elsewhere. 

Thus, the Hamiltonian of the system is   

\begin{equation}
 H=
\begin{pmatrix}
 h(\mathbf{k})+V(\mathbf{r}) & 0\\
 0 & h^*(-\mathbf{k})+V(\mathbf{r})
\end{pmatrix}\,,
\label{Hamiltonian}
\end{equation} 
where $\mathbf k$ is the momentum operator and
\begin{equation}
h(\mathbf{k})=
\begin{pmatrix}
 M\!-\!(B\!+\!D)k^2 & A(k_x+ik_y)\\
 A(k_x-ik_y) & -M\!+\!(B\!-\!D)k^2
\end{pmatrix}\,,
\label{h-Hamiltonian}
\end{equation} 
with $M$, $A$, $B$ and $D$ being the parameters of the BHZ model. The topological phase is realized when $MB>0$~\cite{BHZScience2006,LiuZhang2013inbook}. In practically important cases of the HgTe/CdTe and InAs/GaSb/AlSb quantum wells, the parameters $M, B, D<0$, and $A>0$. The basis set of the wave functions is \mbox{$\{|E_1\uparrow\rangle,|H_1\uparrow\rangle,|E_1\downarrow\rangle,|H_1\downarrow\rangle\}$} where $|E_1\uparrow\rangle$ and $|E_1\downarrow\rangle$ are superpositions of the electron states of $s$-type and the light-hole states of $p$-type with spin up and spin down; $|H_1\uparrow\rangle$ and $|H_1\downarrow \rangle$ are the heavy-hole $p$-type states with opposite spins. In what follows we will restrict ourselves by considering the symmetric model where $D=0$.

As boundary conditions for the wave function, we impose $\Psi(x=0,y)=0$ at the boundary and suppose that the wave function $\Psi(x,y)$ does not diverge at the infinity ($x\to \pm\infty, y\to \infty$). 

A simplification arising from the $S_z$ symmetry of the BHZ model is that the Hamiltonian is block diagonal and therefore it is enough to consider only one spin component.

The eigenfunctions of the Hamiltonian~(\ref{Hamiltonian}) are found using the Fourier transform with respect to $x$ variable and the Laplace transform for $y$ variable. This method allows one to find exactly the eigenfunctions in the case of the short-range potential where the interaction radius is small compared to the characteristic length scale of the wave function. The procedure of the solution was developed previously~\cite{SablikovPRB2015}, so we present here only final equations without technical details.

For convenience we use the dimensionless variables:
\begin{equation}
 \begin{array}{l}
 \varepsilon\!=\!E/|M|,\; \{x',y'\}\!=\!\{x,y\}\sqrt{M/B},\; a\!=\!A/\sqrt{MB},\\
 v(x',y')=V(x,y)/|B|,\;\,\, b=y_d \sqrt{M/B}\,,
 \end{array}
\label{dim_less_param}
\end{equation} 
and in what follows we shall omit the prime in the variables $x, y$.

The Fourier-Laplace transform $\widetilde{\Psi}(k,p)$ of the eigenfunctions is expressed through two variables: $\Phi(k;v,b)$ and $\overline{\Psi}(v,b)$, where $\Phi(k;v,b)$ is the Fourier transform of the $y$-derivative of $\Psi(x,y)$ at the boundary $y=0$ and $\overline{\Psi}(v,b)$ is the value of the wave function at the defect $\overline{\Psi}(v,b)=\Psi(0,b)$. The equation that determines $\widetilde{\Psi}(k,p)$ reads as:
\begin{equation}
 [\varepsilon-h(k,p)]\widetilde{\Psi}(k,p)=\sigma_z\Phi(k;v,b)+\mathrm{I}_{2\times2}\widetilde{v}(k,p)e^{-bp}\overline{\Psi}(v,b)\,.
\label{Fourier-Laplace1}
\end{equation} 
where $\mathrm{I}_{2\times2}$ is the identity $2\times 2$ matrix, $\sigma_z$ is the Pauli matrix, $\widetilde{v}(k,p)$ is the Fourier-Laplace transform of $v(x,y)$. For simplicity, we will present hereafter the equations in a reduced form for $\widetilde{v}(k,p)\approx v$.   

$\overline{\Psi}(v,b)$ is determined by the following equation: 
\begin{equation}
 \left[\mathrm{I}_{2\times2}-\mathcal{K}(\varepsilon;v,b)\right]\overline{\Psi}(v,b)=C(\varepsilon;v,b)\mathcal{F}(\varepsilon;b)\chi,
\label{Psi-bar}
\end{equation} 
where $\mathcal{K}(\varepsilon;v,b)$ and $\mathcal{F}(\varepsilon;v,b)$ are $2\times 2$ matrices which are expressed via the matrix elements of the Hamiltonian $h(\mathbf{k})$; $\chi$ is the following spinor: $\chi=(1,-1)^T$; $C(\varepsilon;v,b)$ is a function, which can be determined by the normalization of the wave function. 

The equations defining $\mathcal{K}(\varepsilon;v,b)$ and $\mathcal{F}(\varepsilon;b)$ are rather cumbersome. In an explicit form they are given in~\ref{Append}.

The spinor $\Phi(k;v,b)$ is determined by the following equation:
\begin{equation}
 \Phi(k;v,b)=-\frac{A'(\varepsilon,k)B(\varepsilon,k;b)}{\Delta_1(\varepsilon,k)}\overline{\Psi}+C(\varepsilon;v,b)\chi\delta[k-k_0(\varepsilon)],
\label{Phi-Psi_1}
\end{equation} 
where $k_0(\varepsilon)$ is the root of the determinant $\Delta_1(\varepsilon,k)$ defined in Eq.~(\ref{Delta1}). It is easy to show from Eqs.~(\ref{Delta1}), (\ref{A}) and (\ref{p12}) that $k_0(\varepsilon)$ is exactly the spectrum of the edge states in the absence of the defect:   
\begin{equation}
 k_0(\varepsilon)=-\frac{\varepsilon}{a}\,.
\label{edge_spectrum}
\end{equation} 

Eqs.~(\ref{Psi-bar}) and~(\ref{Phi-Psi_1}) allow one to calculate the wave function
\begin{multline}
 \Psi(x,y)=\int\limits_{-\infty}^{\infty}\frac{dk}{2\pi}e^{ikx}\int\limits_{c-i\infty}^{c+i\infty}\frac{dp}{2\pi i}\frac{e^{py}} {\Delta(\varepsilon,k,p)}\\
 \times\left[D_0(\varepsilon,k,p)\Phi(k;v,b)+v(k,p)e^{-bp}D_1(\varepsilon,k,p)\overline{\Psi}(v,b)\right].
\label{Psi_Phi_Psi-bar}
\end{multline}

Let us turn to Eq.~(\ref{Psi-bar}) which plays a key role in finding the BIC. This equation has solutions of two kinds depending on the determinant of the matrix in the left-hand side
\begin{multline}
 \Delta_{\Psi}(\varepsilon;v,b)=\bigl[1-\mathcal{K}_{11}(\varepsilon;v,b)\bigr]\bigl[1-\mathcal{K}_{22}(\varepsilon;v,b)\bigr]\\-\mathcal{K}_{12}(\varepsilon;v,b)\mathcal{K}_{21}(\varepsilon;v,b).
\end{multline}

If $\Delta_{\Psi}(\varepsilon;v,b)\ne 0$, the solutions of Eq.~(\ref{Psi-bar}) correspond to the continuum of the edge states flowing around the defect. If the determinant equals zero, this equation has one more solution which exists in the case where the function $C(\varepsilon;v,b)$ is zero. This solution exists only at a discrete value of the energy $\varepsilon_0$ defined by the root of the determinant:
\begin{equation}
\Delta_{\Psi}(\varepsilon_0;v,b)= 0\,.
\label{Delta_Psi}
\end{equation} 
In this case 
\begin{equation}
 \overline{\Psi}(v,b)=C_{bs}\binom{1}{(1-\mathcal{K}_{11})\bigm/\mathcal{K}_{12}}\Biggm|_{\varepsilon=\varepsilon_0}\,,
\label{Psi-bar_BS}
\end{equation} 
with the constant $C_{bs}$ being determined by the normalization. 

For such a solution to exist the wave function should be square integrable. Since $\Psi(x,y)$ is finite and tends to zero at $y\to \infty$, it is sufficient to require that $\Psi(x,y)$ goes to zero when $x\to \pm \infty$ faster than $|x|^{-1/2}$. 

The asymptotics of $\Psi(x,y)$ can be found from Eqs.~(\ref{Psi_Phi_Psi-bar}) and (\ref{Phi-Psi_1}). It has the following form:  
\begin{multline}
 \Psi(x\to\infty,y)\simeq\frac{i e^{ikx}}{8a_{\varepsilon}\frac{\partial \Delta_1}{\partial k}}\left[\frac{e^{-p_2y}}{p_2}D_0(\varepsilon,k,-p_2)\right.\\ 
 -\left.\frac{e^{-p_1y}}{p_1}D_0(\varepsilon,k,-p_1)\right]A'(\varepsilon,k)B(\varepsilon,k;b)\overline{\Psi}(\varepsilon,b)\Biggm|_{k=k_0(\varepsilon)}.
 \label{Psi_asymptotic}
\end{multline}
At this point we will take into account the following property of the matrices $A'(\varepsilon,k)$ and $B(\varepsilon,k;b)$ in the integrand:
\begin{equation}
A'(\varepsilon,k)B(\varepsilon,k;b)\binom{1}{1}\equiv 0\,,
\end{equation} 
which can be verified by direct calculation. Because of this equation we find that $\Psi(x\to \infty,y)\to 0$ if 
\begin{equation}
 \overline{\Psi}=\overline{\psi}\binom{1}{1}.
 \label{Psi-bar_BS1}
\end{equation} 

Combining Eq.~(\ref{Psi-bar_BS1}) with Eqs.~(\ref{Delta_Psi}) and (\ref{Psi-bar_BS}) we arrive at the following equation system that determines a bound state embedded into the continuum of the edge states flowing around the defect:
\begin{equation}
\left\{
\begin{array}{ll}
 1-\mathcal{K}_{11}(\varepsilon;v,b)-\mathcal{K}_{12}(\varepsilon;v,b)&=0\\
 1-\mathcal{K}_{22}(\varepsilon;v,b)-\mathcal{K}_{21}(\varepsilon;v,b)&=0\,.
\end{array}
\right.
\label{BIC_system}
\end{equation} 
A solution of these equations, if it exists, defines the energy $\varepsilon_{bs}$ of the BIC and a relation between $v$ and $b$ under which this state exists. 

\section{Energy of the BIC and conditions of its existence}

In this section we show that Eq.~(\ref{BIC_system}) really has a solution and clarify how the energy of the BIC $\varepsilon_{bs}$ and the potential $v_{bs}$, at which it emerges, depend on the distance $b$ for realistic systems. 

First we prove that Eq.~(\ref{BIC_system}) has one solution when the distance $b$ is large enough. It is convenient to present the matrix $\mathcal{K}(\varepsilon;v,b)$ given by Eq.~(\ref{K}) in the form where the dependence on $v$ is written explicitly
\begin{equation}
 \mathcal{K}(\varepsilon;v,b)=v\left[\mathfrak{K}^{\infty}(\varepsilon)+\mathfrak{K}(\varepsilon;b)\right],
\end{equation}  
Here the first term in the right-hand side is easily shown to be diagonal and the second term exponentially decreases with $b$. 

To begin we consider the limit where $b\to \infty$. In this case $\mathcal{K}_{12}=\mathcal{K}_{21}=0$ and Eq.~(\ref{BIC_system}) takes the form
\begin{equation}
\left\{
\begin{array}{ll}
 1-v\mathfrak{K}^{\infty}_{11}(\varepsilon)&=0\\
 1-v\mathfrak{K}^{\infty}_{22}(\varepsilon)&=0\,.
\end{array}
\right.
\label{BIC_infty}
\end{equation} 
Here the first equation defines the energy of the electronlike bound states, $\varepsilon_e(v)$, and the second equation defines the energy of the holelike bound states, $\varepsilon_h(v)$. The equations are compatible if $\varepsilon_e(v)=\varepsilon_h(v)$. The energies $\varepsilon_e(v)$ and $\varepsilon_h(v)$ as functions of $v$ were investigated in our previous works~\cite{SablikovPSSR2014,SablikovPRB2015}. It was found that the equation $\varepsilon_e(v)=\varepsilon_h(v)$ has a unique solution for both positive and negative $v$. Thus, the equation system~(\ref{BIC_infty}) is satisfied when $\varepsilon=\pm \overline{\varepsilon}$ and correspondingly $v=\pm \overline{v}$.

When the distance $b$ is finite but large enough so that $\mathcal{K}_{12}, \mathcal{K}_{21} \ll 1$, one can seek a solution of Eq.~(\ref{BIC_system}) by expanding $\varepsilon$ and $v$ near $\overline{\varepsilon}$ and $\overline{v}$ at a given $b$. On putting
$\varepsilon=\overline{\varepsilon}+\delta\varepsilon$ and $v=\overline{v}+\delta v$, we arrive at the following equations:
\begin{equation}
\left\{
\begin{array}{ll}
 v^2\mathfrak{K}'_1\delta \varepsilon+\delta v=v^2[\mathfrak{K}_{11}(\overline{\varepsilon};b)-\mathfrak{K}_{12}(\overline{\varepsilon};b)]\\
 v^2\mathfrak{K}'_2\delta \varepsilon+\delta v=v^2[\mathfrak{K}_{22}(\overline{\varepsilon};b)-\mathfrak{K}_{21}(\overline{\varepsilon};b)]\,,
\end{array}
\right.
\label{BIC_b}
\end{equation} 
where $\mathfrak{K}'_{1,2}=\partial_{\varepsilon}\mathfrak{K}^{\infty}_{11,22}(\varepsilon)\big|_{\overline{\varepsilon}}$. It is easy to show that the determinant of this system does not equal zero and there is a unique solution.

A different situation occurs when the distance $b$ is small. In this case the system~(\ref{BIC_system}) turns out to have  no solution. 

Thus the BIC can exist if the distance $b$ exceeds a threshold value and the potential $v$ is equal to a definite value $v_{bs}(b)$ for a given $b$.

\begin{figure}
\centerline{\includegraphics[width=1.\linewidth]{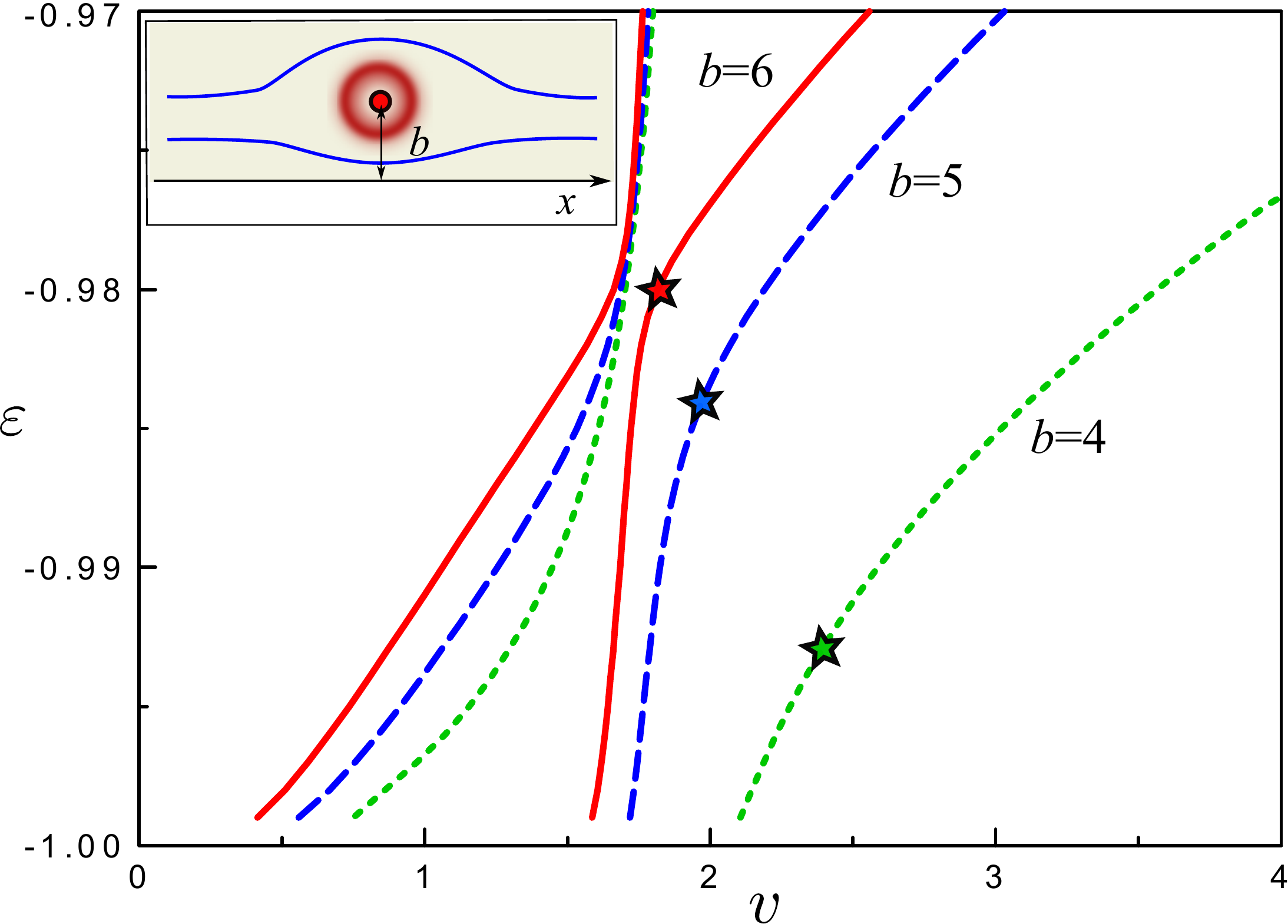}}
\caption{(Color online.) Resonant energies as functions of the defect potential for a variety of distances between the defect and the boundary, $b=4, 5, 6$, near the point where the resonances are degenerate. The asterisks show the energy of the BIC for a given $b$. The calculations were carried out for $a=\sqrt{2}$. The inset illustrates the defect, the BIC and edge states flowing around the defect.}
\label{f_resonances}
\end{figure}

Now we present results of numerical studies of the BIC. Since the BIC arises due to the interference of the resonances in the local density of the edge states flowing around the defect, we present first the resonant energies in the vicinity of the avoided crossing point of the resonances originating from the electronlike and holelike bound states. Fig.~\ref{f_resonances} shows the resonant energies as functions of the defect potential $v$. The points where the BIC arises are shown by the asterisks for different $b$.

\begin{figure}
\centerline{\includegraphics[width=1.\linewidth]{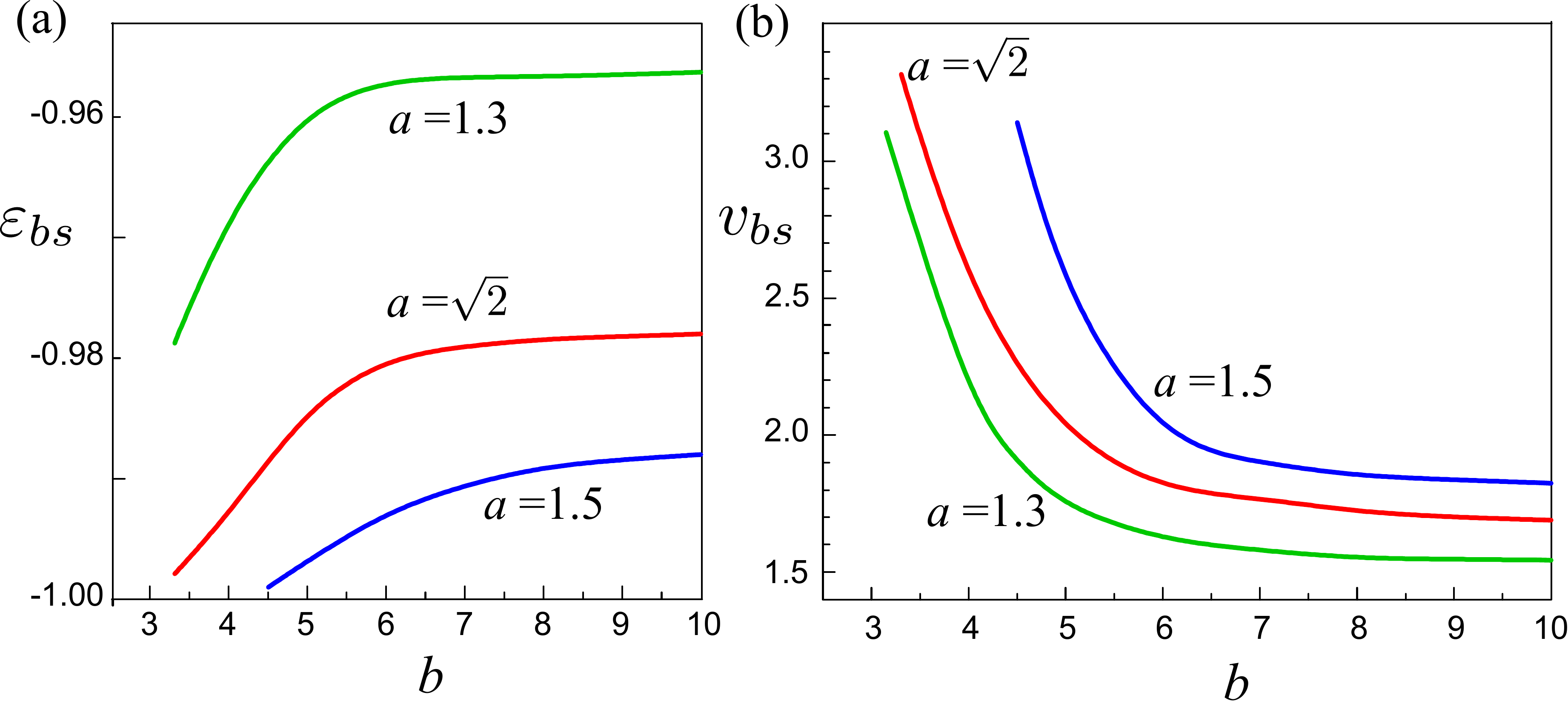}}
\caption{(Color online.) The BIC energy $\varepsilon_{bs}$ (a) and the defect potential $v_{bs}$, at which the BIC arises, (b) as functions of the distance between the defect and the boundary. The calculations were carried out for three values of the parameter $a=1.3, \sqrt{2}, 1.5$.}
\label{f_Ev-b}
\end{figure}

The variation of the BIC energy with the distance between the defect and the boundary is shown in Fig.~\ref{f_Ev-b}(a). Fig.~\ref{f_Ev-b}(b) presents the defect potential $v_{bs}$, at which the BIC arises, as a function of the distance $b$. It is seen that with increasing $b$ the BIC energy and the defect potential tend to the limiting values $\overline{\varepsilon}$ and $\overline{v}$ defined by Eqs.~(\ref{BIC_b}).

\begin{figure}
\centerline{\includegraphics[width=1.\linewidth]{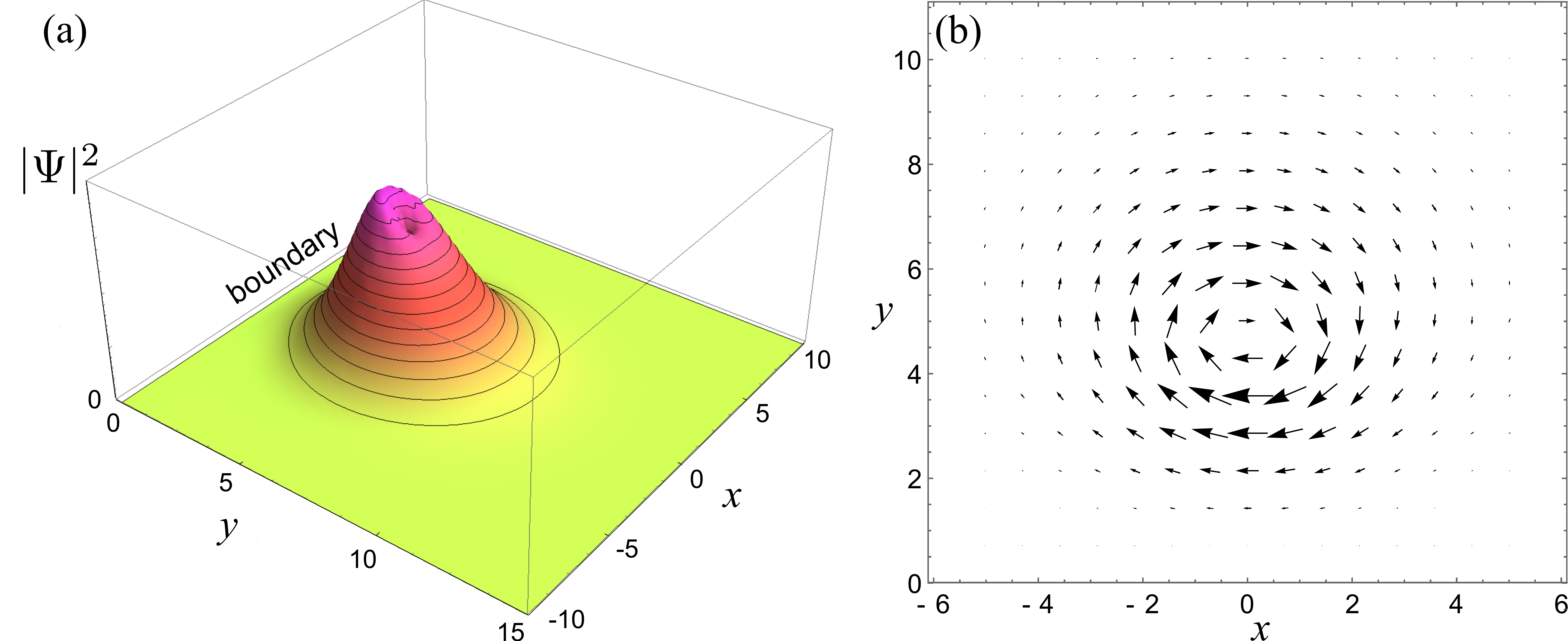}}
\caption{(Color online.) (a) The spacial distribution of the particle density in the BIC induced by the defect located at the distance $b=5$. The BIC energy is $\varepsilon_{bs}=-0.984$. (b) Vector plot of the electron current density in the BIC. The calculations were carried out for the parameter $a=\sqrt{2}$.}
\label{f_Ro_J}
\end{figure}

To complete the description of the BIC in 2D TI we present also the spacial distribution of the particle density in the BIC and the electron current density distribution. The wave function is given explicitly by Eq.~(\ref{Psi_Phi_Psi-bar}) where the matrices in the integrand should be calculated for the energy $\varepsilon_{bs}$ and the defect potential $v_{bs}$. The electron density defined as $|\Psi(x,y)|^2$ is shown in Fig.~\ref{f_Ro_J}(a). The wave function is seen to be strongly localized near the defect. In contrast, the edge states with very close energies extend along the boundary to infinity.

An interesting property of the BIC in 2D TI is the existence of an electron current. The current density can be calculated using the current operator for the BHZ model~\cite{SablikovPRB2015}. Direct calculations show that in the BIC there is a nonzero current whose direction is locked to the spin similarly to the current in the edge states. The vector plot of the current density is shown in Fig.~\ref{f_Ro_J}(b). 

The presence of the circulating current can be interpreted as an indication that the BIC is largely formed by the helical edge modes circulating around the defect. In this regard it should be noted that actually there are two BICs since the state we have found exists for each spin direction. In these states the currents are directed oppositely. However, only one state at a given defect can be occupied by an electron. Therefore around each of the defects, which forms the BIC, there is the electron current, the direction of which is locked to the spin of the trapped electron. 

\section{Conclusion}

The bound states embedded in the energy continuum is a nontrivial quantum phenomenon which attracts increasing interest though they remain fragile objects in spite of the progress in the experiments on nanostructures. Particularly, very special conditions are required for their realization. In this paper we have shown that owing to unique properties of the 2D TIs it becomes possible to create the BICs in the potential of a quite simple form. 

The BIC can arise at a nonmagnetic defect coupled to the edge states in the 2D TIs. The mechanism of the BIC formation is due to the interference of two resonances that exist in the local density of the edge states flowing around the defect. The fact that even the short-range potential of the defect creates two bound states in the bulk of the system is a specific property of 2D TIs, which is due to the presence of two mechanisms of the bound-state formation. In topologically trivial crystals, this property is absent. The resonances are formed as the defect is coupled with the boundary.

To our knowledge such a simple potential inducing the BICs was proposed only in Ref.~\cite{KollarPRL2012} within a one-dimensional two-particle Hubbard model with an attractive impurity potential. The emergence of the BICs on the surface of three-dimensional TIs was noticed in numerical calculations of the electron states in the corner of a ``L''-shaped potential~\cite{SeshadriPRB2014}. 

The system we have considered in this paper favorably differs from above systems since it is quite realizable experimentally. Our calculations are based on the two-band BHZ model that well describes the real systems. Short-range potential is a common minimal model of a defect. Another merit of this model is that it has allowed us to carry out explicit analytical solution of the problem. 

We have found that the BIC can arise if the distance between the defect and the boundary exceeds a threshold value. The BIC emerges at that defect with the definite potential the value of which depends on the distance from the boundary. In the future, it would be interesting to consider also the defect potential of larger radius which creates a larger number of the bound states and correspondingly a larger number of resonances which could facilitate the conditions for the BIC to appear. 

We have shown the electron density in the BIC is localized around the defect and has a dip in the center. We have also found that the electron current flows around the defect. The current direction is locked to the spin which indicates that the BIC has a helical structure.

\section*{Acknowledgments}
This work was partially supported by Russian Foundation for Basic Research (Grant No~14-02-00237) and Russian Academy of Sciences.

\appendix
\section{Matrices $\mathcal{K}(\varepsilon;v,b)$ and $\mathcal{F}(\varepsilon;b)$}
\label{Append}
The explicit equation defining the matrices $\mathcal{K}(\varepsilon;v,b)$ and $\mathcal{F}(\varepsilon;b)$ are as follows:
\begin{multline}
 \mathcal{K}(\varepsilon;v,b)=\int\limits_{-\infty}^{\infty}\frac{dk}{2\pi}\left[\frac{1}{4a_{\varepsilon}\Delta_1(\varepsilon,k)}\mathcal{D}_0(\varepsilon,k)A'(\varepsilon,k)B(\varepsilon,k)\right.\\
 +\biggl.\int\limits_{-i\infty}^{i\infty}\frac{dp}{2\pi i}\frac{v(k,p)}{\Delta(\varepsilon,k,p)}D_1(\varepsilon,k,p)\biggr],
\label{K}
\end{multline}
\begin{equation}
 \mathcal{F}(\varepsilon;b)=\frac{1}{4a_{\varepsilon}}\mathcal{D}_0\bigl(\varepsilon,k_0(\varepsilon);b\bigr),
\end{equation}
where $a_{\varepsilon}=\sqrt{a^2(a^2/4-1)+\varepsilon^2}$ and $\mathcal{D}_0(\varepsilon,k;b)$ denotes the matrix
\begin{equation}
 \mathcal{D}_0(\varepsilon,k;b)\!=\!\frac{e^{-bp_1}}{p_1}D_0(\varepsilon,k,\!-p_1)\!-\!\frac{e^{-bp_2}}{p_2}D_0(\varepsilon,k,\!-p_2).
\end{equation} 
Other matrices, $\mathcal{D}_0, \mathcal{D}_1, A'(\varepsilon,k)$, and $B(\varepsilon,k;b)$, in the integrand are defined as follows:
\begin{equation}
 D_0=
 \begin{pmatrix}
  a_{22}(\varepsilon,k,p) & a_{12}(\varepsilon,k,p)\\
  -a_{21}(\varepsilon,k,p) & -a_{11}(\varepsilon,k,p)
 \end{pmatrix}\,,
\end{equation}
\begin{equation}
D_1=
 \begin{pmatrix}
  a_{22}(\varepsilon,k,p) & -a_{12}(\varepsilon,k,p)\\
  -a_{21}(\varepsilon,k,p) & a_{11}(\varepsilon,k,p)
 \end{pmatrix},
\end{equation} 
\begin{equation}
 A'(\varepsilon,k)=
 \begin{pmatrix}
  a_{12}(\varepsilon,k,p_2) & -a_{12}(\varepsilon,k,p_1)\\
  -a_{22}(\varepsilon,k,p_2) & a_{22}(\varepsilon,k,p_1)
 \end{pmatrix}\,,
\label{A}
\end{equation} 
\begin{multline}
 B(\varepsilon,k;b)=\\
 \begin{pmatrix}
  v(k,p_1)a_{22}(\varepsilon,k,p_1)e^{-bp_1} & -v(k,p_1)a_{12}(\varepsilon,k,p_1)e^{-bp_1}\\
  v(k,p_2)a_{22}(\varepsilon,k,p_2)e^{-bp_2} & -v(k,p_2)a_{12}(\varepsilon,k,p_2)e^{-bp_2}
 \end{pmatrix},
 \label{B}
\end{multline}
where $a_{ij}(\varepsilon,k,p)$ are the elements of the matrix $[\varepsilon-h(k,p)]$:
\begin{equation}
 \begin{array}{ll}
 a_{11}=\varepsilon+1-k^2+p^2, & a_{12}=-a(k+p),\\
 a_{21}=-a(k-p), & a_{22}=\varepsilon-1+k^2-p^2.
 \end{array}
\end{equation} 

In addition, the scalar functions $\Delta_1(\varepsilon,k), \Delta(\varepsilon,k,p)$, and $p_{1,2}(\varepsilon,k)$ have the form:
\begin{multline}
  \Delta_1(\varepsilon,k)=a_{22}(\varepsilon,k,p_1)a_{12}(\varepsilon,k,p_2)\\-a_{12}(\varepsilon,k,p_1)a_{22}(\varepsilon,k,p_2)\,, 
\label{Delta1}
\end{multline}
\begin{equation}
 \Delta(\varepsilon,k,p)=-\left[p^2-p_1^2(\varepsilon,k)\right]\left[p^2-p_2^2(\varepsilon,k)\right],
\end{equation} 
with
\begin{equation}
 p_{1,2}(\varepsilon,k)=\sqrt{k^2+a^2/2-1\pm\sqrt{a^2(a^2-4)/4+\varepsilon^2}}
 \label{p12}
\end{equation} 
and $\mathrm{Re}p_{1,2}(\varepsilon,k)\ge 0$.
\bibliography{BIC_2DTI_c}
\end{document}